\documentclass[pre,aps,amsmath,amssymb,floatfix,twocolumn,longbibliography]{revtex4-1}

\pdfoutput=1
\usepackage{siunitx}
\usepackage{graphicx}
\usepackage{listings}
\usepackage[table,xcdraw]{xcolor}
\usepackage{bm}
\usepackage{hyperref}

\usepackage{array} 
\newcolumntype{C}[1]{>{\centering\arraybackslash}p{#1}} 

\newcommand{\code}[1]{\texttt{\detokenize{#1}}}

\def\eqq#1{Eq.~(\ref{#1})}

\def\eq#1{(\ref{#1})}

\def\f#1{Fig.~\ref{#1}}

\def\s#1{Section~\ref{#1}}

\def\c#1{~\cite{#1}}

\def\cc#1{~Ref.~\cite{#1}}
\def\ccc#1{~Refs.~\cite{#1}}

\def\av#1{\langle #1 \rangle}

\def\beq{\begin{equation}}
\def\eeq{\end{equation}}
\def\bea{\begin{eqnarray}}
\def\eea{\end{eqnarray}}

\def\cee{{\bm c}}

\def\kt{k_{\rm B}T}
\def\tf{t_{\rm f}}

\begin{document}

\title{Benchmark control problems in nonequilibrium statistical mechanics}
\author{Stephen Whitelam}\email{swhitelam@lbl.gov}
\affiliation{Molecular Foundry, Lawrence Berkeley National Laboratory, 1 Cyclotron Road, Berkeley, CA 94720, USA}
\author{Corneel Casert}
\email{ccasert@lbl.gov }
\affiliation{NERSC, Lawrence Berkeley National Laboratory, 1 Cyclotron Road, Berkeley, CA 94720, USA}
\author{Megan Engel}
\email{megan.engel@ucalgary.ca}
\affiliation{Department of Biological Sciences, University of Calgary, Calgary, AB, Canada}
\author{Isaac Tamblyn}
\email{itamblyn@block.xyz}
\affiliation{Block, Toronto, ON M5A 1J7, Canada}
\affiliation{Department of Physics, University of Ottawa, Ottawa, ON K1N 6N5, Canada}
\affiliation {Vector Institute for Artificial Intelligence, Toronto, ON M5G 1M1, Canada}

\begin{abstract}
We present a set of computer codes designed to test methods for optimizing time-dependent control protocols in fluctuating nonequilibrium systems. Each problem consists of a stochastic model, an optimization objective, and C++ and Python implementations that can be run on Unix-like systems. These benchmark systems are simple enough to run on a laptop, but challenging enough to test the capabilities of modern optimization methods. This release includes five problems and a worked example. The problem set is called NESTbench25, for NonEquilibrium STatistical mechanics benchmarks (2025).
\end{abstract}
\maketitle

\section{Introduction}

Recent interest in finding optimal time-dependent protocols for stochastic dynamics has been motivated by the need to control strongly fluctuating experiments and the desire to explore the results of stochastic thermodynamics, including fluctuation relations, speed limits, and uncertainty relations \c{seifert2012stochastic,proesmans2020optimal,blaber2023optimal,ciliberto2017experiments,alvarado2025optimal}. A variety of numerical approaches have been developed for these purposes, including methods for evaluating thermodynamic geodesics\c{rotskoff2015optimal,sivak2012thermodynamic,zhong2024beyond}, path-sampling techniques\c{gingrich2016near}, variational methods\c{das2023nonequilibrium}, Monte Carlo schemes\c{werner2024optimized}, evolutionary strategies \c{whitelam2023demon}, and gradient-based algorithms\c{geiger2010optimum,rengifo2025machine,engel2023optimal}.

In this paper we introduce a set of benchmark problems and associated computer codes designed to enable users to test methods designed to optimize time-dependent protocols. This set of problems, motivated in part by the widespread use of benchmarks in machine learning, offers various tests of an optimization method, and should be solvable, or nearly so, by a variety of methods.

The problems in this set were introduced by other authors, and we have selected them for their physical interest and varied challenges. We consider overdamped\c{schmiedl2007optimal} and underdamped\c{gomez2008optimal} Langevin particles in a moving harmonic trap; information erasure using an underdamped Langevin particle\c{dago2021information,Dago-2024-APR}; an active Brownian particle confined in a trap\c{baldovin2023control}; and magnetization reversal in the Ising model\c{rotskoff2015optimal}. In the trap problems the aim is to translate the trap with least work. In the erasure problem the aim is to perform erasure as reliably as possible. The active particle must be brought from a passive equilibrium state to a designated active steady state, while extracting from the active bath as much work as possible. The state of the Ising model must be changed with as little entropy produced as possible. All tasks must be accomplished in finite time $\tf$, which is chosen to be short relative to the timescales of each system. This choice challenges analytic methods that identify efficient protocols in the long-time limit\c{sivak2012thermodynamic}, and those analytic methods in turn provide valuable bounds with which to assess the effectiveness of numerical methods. 

The physical complexity of the systems increases steadily throughout the set, moving from a one-body system with one control parameter to a many-body system with two control parameters. However, this increase of system complexity does not necessarily correspond to an increase in the difficulty of controlling these systems as efficiently as possible. We have chosen problems whose difficulty is roughly comparable to that of MNIST\c{mnist} in machine learning: problems are simple enough to run on a laptop, accessible to a number of methods, but contain enough complexity to provide a non-trivial test of a control algorithm. One point to note: the success of a protocol is determined by its performance with respect to an order parameter of interest. For many problems it is possible to observe protocols that are visually quite different, but that perform similarly.

Some of these systems can be recast in deterministic terms, via Fokker-Planck equations, and solved analytically. However, we present the fluctuating version of each system: our intent is to understand how control algorithms cope with noisy trajectories, and so might cope with a complex system for which a deterministic description is not feasible. We have chosen protocols parametrized in terms of time alone (sometimes called closed-loop or feedforward protocols) rather than protocols parameterized also as a function of the system's microscopic state (sometimes called open-loop or feedback-control protocols), which we save for a future release. This release we call NESTbench25, for NonEquilibrium STatistical mechanics (or alternatively Stochastic Thermodynamics) benchmarks (2025).

In \s{systems} we describe the problems, summarized in Table~\ref{tab1}, and explain the structure of the corresponding codes used to run them. The codes are implemented in C++ and Python (the figures in the text were produced using the C++ codes). The C++ codes use mostly C-style logic, with limited use of C++ for certain libraries and input/output, and are written in a simple procedural style. The Python code uses the PyTorch framework, allowing for parallel computation on both GPU and CPU.

In \s{worked_example} we provide a worked example of protocol learning, providing C++ and Python implementations of a neurevolutionary learning algorithm for the overdamped trap-translation problem, and a JAX implementation of automatic differentiation for the same. We also provide the results of neurevolutionary learning for the other problems, and discuss some of the challenges associated with them.

\begin{table*}[]
\renewcommand{\arraystretch}{1.3} 
\centering
\begin{tabular}{|C{3.2cm}|C{4.2cm}|C{3.6cm}|C{4cm}|C{2cm}|}
\hline
\textbf{Problem} & \textbf{Description} & \textbf{Objective} & \textbf{Control parameters $\cee(t)$} & \textbf{Analytic solution}? \\
\hline
\code{trap_overdamped} & Overdamped Langevin particle in a moving harmonic trap & Minimize work & Trap center $\lambda$ &Yes \\
\code{trap_underdamped} & Underdamped Langevin particle in a moving harmonic trap & Minimize work & Trap center $\lambda$&Yes \\
\texttt{erasure} & Logic erasure via underdamped dynamics & Minimize error rate & Potential shape $c_0$ and position $c_1$ &No \\
\texttt{abp} & Active Brownian particle in a harmonic trap & Enact state-to-state transformation and maximize extracted work & Trap stiffness $\kappa$ and swim speed $\lambda$ & No \\
\texttt{ising} & 2D Ising model undergoing state change & Minimize entropy production & Temperature $T$ and magnetic field $h$ & No \\
\hline
\end{tabular}
\caption{Summary of the problems included in this benchmark set.}
\label{tab1}
\end{table*}

The codes are available at the GitHub repository \href{https://github.com/protocol-benchmarks/NESTbench25}{\texttt{protocol-benchmarks/NESTbench25}}, in the folders \code{c_code}, \code{python_code}, and \code{jax_code}. This paper serves as the \code{README} file for this set of codes. Comments, suggestions, portability issues, and bug reports are welcome: please use the email addresses listed.

\section{The problems and their implementations}
\label{systems}

\subsection{Translating an overdamped particle with minimum work}
\label{trap_overdamped_1}

The code \code{engine_trap_overdamped} (\code{.cpp} or \code{.py}) in the folder \code{trap_overdamped} simulates the first problem of\cc{schmiedl2007optimal}, the translation of an overdamped particle by a harmonic trap. We consider a point particle at position $x$ in a potential
\beq
\label{you}
U(x,\lambda)=\frac{1}{2} \left(x-\lambda(t) \right)^2,
\eeq 
in units such that $\kt=1$. In general, we denote the vector of control parameters of the problem as $\cee(t) = (c_0(t),c_1(t),\dots)$; here, the trap center $c_0(t)=\lambda(t)$ is sole control parameter of the problem.

The particle undergoes the Langevin dynamics
\beq
\label{langevin}
\gamma \dot{x}=-\frac{\partial U(x,\lambda(t))}{\partial x}+ \left(2 \gamma \kt\right)^{1/2}\xi(t),
\eeq
where dots denote differentiation with respect to time $t$, and the Gaussian white noise $\xi$ has zero mean and correlation function $\av{\xi(t) \xi(t')} = \delta(t-t')$. We set the damping coefficient $\gamma$ to 1, and work in units such that $\kt=1$.

The aim of this problem is to move the trap center from an initial position $\lambda(0)=0$ to a final position $\lambda(\tf)=5$, in finite time $t_{\rm f}=1$, minimizing the work 
\beq
\label{eq_work}
\av{W} = \int_0^{\tf} {\rm d}t\, \dot{c}_0(t) \left\langle \frac{\partial U(x, \lambda)}{\partial \lambda} \right\rangle\eeq
averaged over many realizations of the process. The optimal (mean-work-minimizing) protocol is known analytically\c{schmiedl2007optimal}. It has a linear form 
\beq
\lambda^\star(t)=\lambda(\tf) \frac{t+1}{t_{\rm f}+2}
\eeq
 for $0<t<t_{\rm f}$, and has jump discontinuities at the start $(t=0)$ and end $(t=t_{\rm f})$; see \f{fig_trap_overdamped_1}(a). This protocol produces mean work $\av{W}^\star=\lambda(\tf)^2/(t_{\rm f}+2) =25/3\approx 8.333$\c{schmiedl2007optimal}. 

The jump discontinuities are a generic feature of many optimal protocols\c{blaber2021steps}. To learn them requires a protocol ansatz that is not constrained to smoothly approach its boundary values, which this problem tests. Otherwise, learning a linear profile in the temporal bulk $0< t < \tf$ is a straightforward task, and has been done by several numerical approaches\c{geiger2010optimum,rengifo2025machine,whitelam2023demon,engel2023optimal}.

In the code \code{engine_trap_overdamped} the user specifies  the time-dependent protocol $0 < \lambda(t) < \tf$ via the external file \code{input_control_parameters.dat}, whose dimensions are specified by the function \code{load_protocol()}. The external file sets the internal lookup table \code{cee} (which stands in general for $\cee(t)$, the control-parameter vector); the boundary conditions at $t=0$ and $t=\tf$ are set in the code by \code{cee_boundary}, which the user does not need to specify.

For each example we also include a default protocol, which can be called from \code{load_default_protocol()}. For this problem the default protocol is the optimal protocol of\cc{schmiedl2007optimal}. 
\begin{figure}[]
   \centering
   \includegraphics[width=\linewidth]{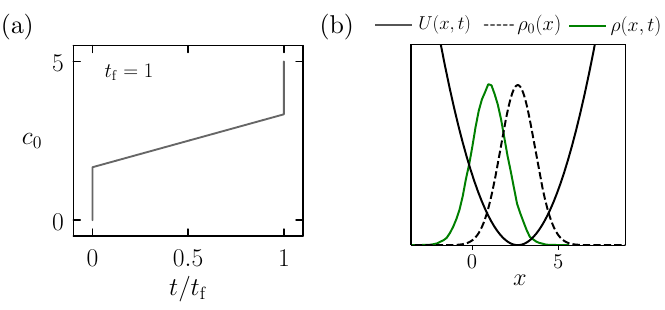} 
   \caption{Translation of an overdamped particle by a harmonic trap, taken from\cc{schmiedl2007optimal}. The problem is to specify the time-dependent protocol $c_0(t)$, where $c_0=\lambda$ is the trap center, in order to minimize the mean work done upon many translations. (a) Optimal protocol known from analytic calculation\c{schmiedl2007optimal}; this is the default protocol called from \code{load_default_protocol()}. (b) Snapshot (mid-trajectory) produced by the function \code{visualize_protocol()}. Here $U(x,\lambda(t))$ is the potential, $\rho_0(x)$ is the associated Boltzmann distribution, and $\rho(x,t)$ is the distribution of particle positions.}
   \label{fig_trap_overdamped_1}
\end{figure}

An illustration of how this protocol works is provided by the \code{main()} function in the C++ code (the other codes are structured similarly). To compile and run the C++ engine as stand-alone code, do \code{make standalone} (ensure \code{Makefile} is in the same directory as the code). The Python version of this code will perform the calculations on a GPU if available.

The \code{main()} function seeds the random number generator, loads the default protocol, and outputs both the current and optimal protocols so that you can check (in this case) that they are the same. To load your own protocol, replace \code{load_default_protocol()} with \code{load_protocol()}, making sure that you have provided the file \code{input_control_parameters.dat}. The dimensions of the latter file is $C \times 1000$, where $C=1$ (or 2, in later sections) is the number of control parameters of the problem, and 1000 the number of time slices at which the protocol is changed. Each folder also contains \code{input_control_parameters_learned.dat} (obtained as described in \s{worked_example}), which can be loaded by changing the name of that file to \code{input_control_parameters.dat}.

The code integrates \eqq{langevin} using the first-order Euler scheme
\beq
x(t+\Delta t) = x(t)-\partial_x U(x,\lambda(t)) \Delta t + \sqrt{2 \Delta t}X,
\eeq
where $\Delta t = 10^{-3}$ and $X$ is a Gaussian random number with zero mean and unit variance. 

The function \code{visualize_protocol()} calculates, using $10^5$ independent trajectories, the mean work $\av{W}$. It also outputs a histogram of work, \code{report_work_histogram.dat}, and outputs a picture \code{output_figure.pdf} and a movie \code{output_movie.mp4} showing time-ordered snapshots of the probability distribution of $x$. One frame of this movie is shown in \f{fig_trap_overdamped_1}(b). (For the C++ version, make sure the movie- and picture maker \code{movie.py} is in the same directory as the simulation engine.)

The function \code{final_answer()} calculates the order parameter for the problem, in this case the mean work $\av{W}$, over 100 samples of $10^4$ trajectories. Under the default (optimal) protocol we obtain \code{order parameter =} $8.32989 \pm 0.00468$, which agrees with the analytic value of $25/3$ to within statistical error.

In \s{worked_example} we illustrate how to use the code as an external library in order to do protocol learning.

\subsection {Translating an underdamped particle with minimum work}
\label{trap_underdamped_1}

The code \code{engine_trap_underdamped} (\code{.cpp} or \code{.py}) in the folder \code{trap_underdamped} simulates the underdamped analog of the previous problem, taken from\cc{gomez2008optimal}. In this problem, a particle with position $x$ experiences the underdamped Langevin dynamics
\beq
\label{lang2}
m \ddot{x}+ \gamma \dot{x} = -\frac{\partial U(x,\lambda(t))}{\partial x}+\left( 2 \gamma \kt\right)^{1/2}\xi(t).
\eeq
Here dots denote differentiation with respect to time $t$, $m$ is the particle mass and $\gamma$ the damping coefficient, and the Gaussian white noise satisfies $\av{\xi(t)}=0$ and $\av{\xi(t)\xi(t')}=\delta(t-t')$. The function $U(x,\lambda)$ is a harmonic potential given by
\beq
\label{you2}
U(x,\lambda)=\frac{k}{2} \left(x-\lambda(t) \right)^2,
\eeq 
where $k$ is the spring constant. As before, the trap center $c_0(t)=\lambda(t)$ is the control parameter of the problem, set for $0 < t < \tf$ by the file \code{input_control_parameters.dat} (which the user must specify). The boundary conditions impose the values $\lambda(0)=0$ and $\lambda(\tf)=5$. As before, the order parameter to be minimized is the mean work required to perform many independent realizations of this translation.  

The natural scales of this problem are $\sigma_0 = \sqrt{\kt/k}$ for length, $\omega_0^{-1} = \sqrt{m/k}$ for time, and $\kt$ for energy, and it will be convenient to introduce the quality factor $Q \equiv m \omega_0/\gamma$ of a harmonic oscillator~\footnote{Note that we can rescale space as $x \to \sigma_0^{-1} x$, time as $t \to \omega_0 t$, and the potential as $U \to U/\kt$ in order to write \eqq{lang2} as
\beq
\label{lang2b}
\ddot{x}+ Q^{-1} \dot{x} = -\partial_x U(x,\lambda(t))+\sqrt{ 2 Q} \,\xi(t),
\eeq
with all parameters dimensionless.}. The values of $\sigma_0 = 1$ nm, $Q=7$, and $\omega_0= 2 \pi \times 1090$ Hz chosen in \code{engine_trap_underdamped} are motivated by the values of the micromechanical cantilever of\ccc{Dago-2021,dago2021information}, and so the `particle' in this problem represents the position of a cantilever tip. The protocol time is set to be $\tf=0.15 \times (2 \pi/\omega_0)$, short relative to the fundamental oscillation time $2 \pi/\omega_0$. 

The code \code{engine_trap_underdamped} integrates \eqq{lang2} using an Euler-like scheme. Introducing the velocity $v=\dot{x}$ and integrating \eq{lang2} over the short time interval $\Delta t$ yields the update equations
\beq
\label{ex}
x(t+\Delta t)=x(t)+v(t) \Delta t,
\eeq
and
\bea
\label{vee}
v(t+\Delta t)&=& \lambda v(t)- \left(1-\lambda\right) \gamma^{-1}\partial_x U(x,\lambda) \nonumber \\&+&  \sqrt{1-\lambda^2}\, \Omega X,
\eea
where $\lambda \equiv {\rm e}^{-\gamma \Delta t/m}$, $\Omega \equiv \sqrt{ k_{\rm B} T/m }=\sigma_0 \omega_0$, and $X$ is a Gaussian random number of zero mean and unit variance. The parameter combinations that appear in \eq{vee} are $\gamma/m=\omega_0/Q$ (in the term $\lambda$), $k\sigma_0/\gamma=Q \omega_0 \sigma_0$ (in the term in $\partial_x U(x,\lambda)$), and $\Omega=\omega_0 \sigma_0$ (in the noise term)~\footnote{Expressing time and length in dimensionless units, \eqq{vee} is equivalent to
\beq
\label{vee2}
v(t+\Delta t)= \lambda v(t)- \left(1-\lambda\right) Q\partial_x U(x,\lambda) +  \sqrt{1-\lambda^2}\, X,
\eeq
with $\lambda \equiv {\rm e}^{-\Delta t/Q}$.}. 
\begin{figure}[]
   \centering
   \includegraphics[width=\linewidth]{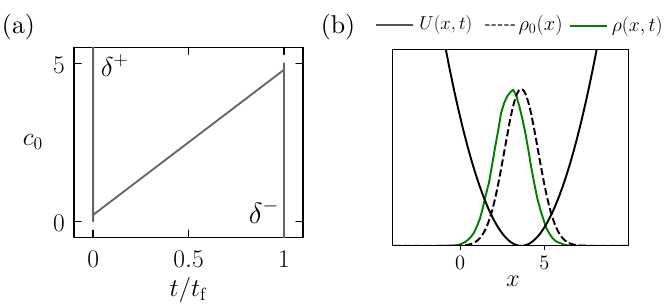} 
   \caption{Translation of an underdamped particle by a harmonic trap, taken from\cc{gomez2008optimal}. (a) Optimal protocol known from analytic calculation; this is the default protocol called from \code{load_default_protocol()}. (b) Snapshot produced by the function \code{visualize_protocol()}. Here $U(x,\lambda(t))$ is the potential, $\rho_0(x)$ is the associated Boltzmann distribution, and $\rho(x,t)$ is the distribution of particle positions.}
   \label{fig_trap_underdamped_1}
\end{figure}
\begin{figure*}[]
   \centering
   \includegraphics[width=0.75\linewidth]{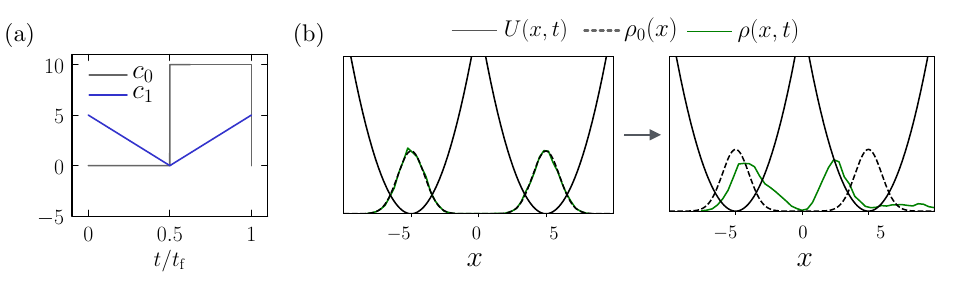} 
   \caption{Logic erasure modeled by an underdamped particle in the potential \eq{pot}. (a) The default protocol for the two control parameters of the potential. (b) For the chosen erasure time, this protocol is ineffective, producing an erasure rate of little more than 50\%: the snapshot shows that about half of the particle-position distribution fails to reach the left-hand sector. Here $U(x,\cee(t))$ is the potential, $\rho_0(x)$ is the associated Boltzmann distribution, and $\rho(x,t)$ the distribution of particle positions.}
   \label{fig_erasure_1}
\end{figure*}

The protocol $\lambda^\star(t)$ that moves the trap from position $\lambda(0)=0$ to position $\lambda(\tf)$ in time $t_{\rm f}$ with minimum work is known analytically, and is given by
\beq
\label{opt}
\lambda^\star(t)=\lambda(\tf)\frac{Q\omega_0 t+1}{Q\omega_0 t_{\rm f}+2}+\Lambda \left[ \delta(t) - \delta(t-t_{\rm f})\right],
\eeq
where $\Lambda \equiv \lambda(\tf)/(\omega_0^2 t_{\rm f}+2 \omega_0 Q^{-1})$. This protocol is sketched in \f{fig_trap_underdamped_1}(a).

The optimal protocol for this problem contains delta-function kicks, which effect an instantaneous change of the velocity of the particle. This poses a challenge for an optimization algorithm, because the protocol does not directly control the velocity of the particle, only the trap position $\lambda(t)$. Part of the challenge of this problem is to find a protocol that approximates the effect of a delta function impulse. In addition, we must ensure that the timestep used in the code is small enough to accommodate this protocol without breaking the integrator. This issue is discussed at more length in \s{trap_underdamped_2}.

The mean work done under this protocol is
\beq
\label{opt_work}
W^\star=\frac{k \lambda(\tf)^2}{2+Q\omega_0t_{\rm f}}.
\eeq

The function \code{default_protocol()} calls the optimal protocol \eq{opt}. The first term in \eq{opt} is encoded by the trap position lookup table \code{cee}, while the second term is imposed directly on the velocity of the particle via the terms \code{delta_vee}: in the presence of the delta-function impulses, \eqq{vee} acquires an extra term
\beq
\Delta v \left( \delta_{t,0}-\delta_{t,t_{\rm f}}\right),
\eeq
where $\Delta v \equiv k \Lambda/m = \lambda(\tf)/(t_{\rm f}+2 Q^{-1} \omega_0^{-1})$. The energy change associated with the delta kicks (which we count as work) is $\Delta W = m \Delta v(v_{\rm i}-v_{\rm f})$, where $v_{\rm i}$ is the velocity immediately before the first kick (at $t=0$) and $v_{\rm f}$ is the velocity immediately after the second kick (at $t=\tf$). Note that $m=k_{\rm B}T/(\sigma_0^2 \omega_0^2)$.

With the chosen model parameters and simulation time $\tf$, the optimal work value \eq{opt_work} is 
\beq
W^\star = \frac{1 \times 5^2}{2+7 \times 2 \pi \times 0.15} \approx 2.90787,
\eeq
 in units of $\kt$. For our numerical implementation of the default (optimal) protocol, the function \code{final_answer()} returns an estimate of the mean work as \code{order parameter} $= 2.90415 \pm 0.0027$, using 100 samples of $10^4$ trajectories. 
 
The function \code{visualize_protocol()} calculates the mean work and work distribution, and outputs a movie and set of time-ordered snapshots illustrating the action of the protocol. One such snapshot is shown in \f{fig_trap_underdamped_1}(b).

\subsection{Maximizing the probability of logic erasure using an underdamped particle}
\label{erasure_1}

The code \code{engine_erasure} in the folder \code{erasure} simulates the classic physics- and information theory problem of logic erasure\c{landauer1961irreversibility}. Unlike for the systems of \s{trap_overdamped_1} and \s{trap_underdamped_1}, there is no known analytic solution to this optimization problem.

 An underdamped Langevin particle can start in either well of a double-well potential, a physicist's model of a one-bit computer memory. The left- and right-hand wells of the potential are associated with logic states 0 and 1, respectively. We are allowed to manipulate two parameters of the potential as a function of time, with the aim of bringing the particle to the left-hand well at time $\tf$. If successful, this process erases information, because observing the final state of the system provides no information as to its starting state. Erasing a one-bit memory costs at least $\kt \ln 2$ units of energy\c{landauer1961irreversibility}, and additional energy is required as the rate of erasure increases\c{Proesmans-2020,Dago-2021}. In this problem, erasure must be performed so quickly that errors are inevitable. The order parameter to be minimized is the error rate of erasure.

In detail, the model is based on the micromechanical cantilever of\cc{dago2021information,Dago-2022-JStat}. A particle, modeling the position of the cantilever tip, experiences the underdamped Langevin dynamics
\beq
\label{lang3}
m \ddot{x}+ \gamma \dot{x} = -\frac{\partial U(x,\cee(t))}{\partial x}+\left( 2 \gamma \kt\right)^{1/2}\xi(t).
\eeq
The variables of this equation are as in \s{trap_underdamped_1}, with the exception of the potential 
\bea
\label{pot1}
U(x, \mathbf{c}(t)) &=& \frac{k}{2} \left( x - S(x - c_0(t))\, c_1(t) \right)^2 \\
&+& k c_0(t)\, c_1(t) \left( S(x - c_0(t)) + S(c_0(t)) \right). \nonumber
\label{pot}
\eea
Here $S$ is the sign function, where $S(x) = - 1$ (resp. 1) if $x < 0$ (resp $x > 0$). The fundamental lengthscale, timescale, and quality factor of the oscillator, $\sigma_0 = 1$ nm, $\omega_0= 2 \pi \times 1090$ Hz, and $Q=7$, are as in \s{trap_underdamped_1}. The integration scheme in the code \code{engine_erasure} is the same as in \s{trap_underdamped_1}.

The potential \eq{pot1} has in general a double-well form, shown in \f{fig_erasure_1}. It is parameterized by $\cee(t) = (c_0(t), c_1(t))$, which are the two control parameters of our problem. The starting and ending values of these parameters are $\cee(0)=\cee(\tf)=(0,5)$, giving a double-well potential with an energy barrier of $12.5$ (in units of $\kt$). The memory is therefore volatile: thermal fluctuations can drive the particle over the barrier.

 Starting in equilibrium with this potential, with the particle in either well with equal likelihood, erasure is deemed successful if $x(\tf)<0$. The difficulty of this problem comes from the timescale under which erasure must be performed. The protocol time is set to be $\tf=0.5 \times 2 \pi/\omega_0$, smaller than the fundamental oscillation time of the particle, and half the time considered in\cc{barros2025learning}. Large forces must be employed in order to effect erasure on short timescales, but these forces impart a large kinetic energy to the particle, potentially allowing it to escape the intended potential well and so causing erasure to fail. For this reason this problem includes a quiescent period (during which the potential is held in its default double-well form) of time $2\tf$ after the protocol has run, in order to check the stability of the memory. The protocol specified by the user in \code{input_control_parameters.dat} affects only the initial period of time $\tf$.

The choice of \code{default_protocol()} illustrates this overheating problem. This protocol, shown in \f{fig_erasure_1}(a), is an intuitive one that merges the double wells, translates the resulting single well to the left, and then reconstitutes the double-well potential. For sufficiently long erasure times, several multiples of $2 \pi/\omega_0$, the erasure probability under this protocol is close to unity. But for erasure time $\tf=0.5 \times 2 \pi/\omega_0$, the probability of successful erasure is only about 53\%. The movie generated by \code{visualize_protocol()} shows the problem: the kinetic energy imparted to the particle is so large that its position-probability distribution oscillates back and forth, jumping over the energy barrier after the latter is reconstituted. The snapshot in \f{fig_erasure_1}(b) shows the final frame of the movie. 

The function \code{final_answer()} computes the erasure failure rate over 100 samples of $10^4$ trajectories, and returns \code{order parameter} $= 0.477974 \pm 0.000456$ for the default trajectory. The goal of this problem is to find a protocol that makes this order parameter as small as possible.

\subsection{Enacting a state-to-state transformation and extracting work from an active Brownian particle in a trap}
\label{abp_1}

The code \code{engine_abp} in the folder \code{abp} simulates a modification of the problem of\cc{baldovin2023control}, involving an active Brownian particle in two-dimensional space. The goal of this problem is to bring a distribution of noninteracting active particles (calculated using many independent simulations of a single particle) from a passive state to an active state {\em and} to extract from the system as much work as possible. Energy extraction is possible because we do not account for the energy that powers the active particle: some of this energy can be transferred to the trap and recorded as work extracted from the system.

 In reduced units, the particle has position vector $\bm{r}$ and orientation $\theta$, and moves in the direction $ \hat{\bm{e}}(\theta)=(\cos\theta, \sin\theta)$ with constant speed $\lambda$. Its motion is governed by the Langevin equations 
\begin{align}
\label{lang_abp}
    \begin{split}
   \dot{{\bm r}} &= \lambda(t) \hat{\bm{e}}(\theta) - \kappa(t) \bm{r} + \sqrt{2}\bm{\xi}_r(t),\\ 
\dot{\theta} & = \sqrt{2}\xi_\theta(t),
    \end{split}
\end{align}
where $\bm{\xi}_r(\tau)$ and $\xi_\theta(\tau)$ are Gaussian white noise terms with zero mean and unit variance. The force term $-\kappa(t) {\bm r}$ is derived from a harmonic restoring potential $U({\bm r},\kappa(t))=\frac{1}{2} \kappa(t) {\bm r}^2$. The spring constant $\kappa(t)$ and swim speed $\lambda(t)$ are the control parameters $\cee(t)$ of the problem. 

This model is inspired by experiments on spherical Janus particles, whose self-propulsion speed is controlled by light intensity, and which are confined by acoustic traps. In typical setups, control parameters are bounded as $0 \le \lambda \le 11$ and $1 \le \kappa \le 7$\c{baldovin2023control,buttinoni2012active,takatori2016acoustic}. These bounds are enforced by the code, and do not need to be enforced in the file \code{input_control_parameters.dat}, which is specified by the user.

The problem studied in\cc{baldovin2023control} is to find a time-dependent protocol $\cee(t) = (c_0(t),c_1(t))=(\kappa(t),\lambda(t))$ that obeys the bounds of the previous paragraph and that minimizes the time $\tf$ required to take the system from a passive steady state at $(\kappa(0),\lambda(0)) = (4,2.5)$ to an active one at $(\kappa(\tf),\lambda(\tf)) = (4,5)$. Using a constrained ansatz in order to make the problem analytically tractable, the transformation time found in \cc{baldovin2023control} was $\tf \approx 0.44$. Using numerics allows us to remove these constraints, which allows the transformation to be done faster, in time $\tf \approx 0.15$\c{casert2024learning}. 

\begin{figure}[]
   \centering
   \includegraphics[width=\linewidth]{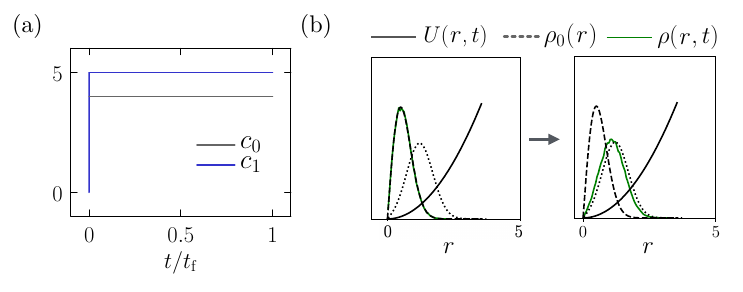} 
   \caption{Active Brownian particle in a trap, modified from\ccc{baldovin2023control,casert2024learning}. The goal of this problem is to induce a state-to-state transformation while extracting as much work as possible. (a) The default protocol for the two control parameters ${\bm c(t)} = (\kappa(t),\lambda(t))$: the trap spring constant $\kappa$ and the swim speed $\lambda$ are set equal to their final-time values (and so $\kappa$ is unchanged). (b) Snapshot created by \code{visualize_protocol()} illustrating the action of the default protocol. Here $U(r,{\bm c}(t))$ is the harmonic potential in which the particle is confined, $\rho_0(r)$ is the associated Boltzmann distribution, and $\rho(r,t)$ the distribution of particle positions. The target distribution is the outer dotted profile: the default protocol fails to enact a state-to-state transformation.}
   \label{fig_abp_1}
\end{figure}

Here we consider a modification of this problem, following\cc{casert2024learning}. We start in the passive limit $(\kappa(0),\lambda(0)) = (4,0)$, so that the initial distribution is a Boltzmann one, and end with the parameters $(\kappa(\tf),\lambda(\tf)) = (4,5)$. As in the other problem, the initial- and final-time boundary values are enforced by the code. At time $\tf$ we require that the particle radial distribution $\rho(r,\tf)$ is consistent with the steady-state value $\rho_{\rm ss}(r)$ associated with the parameters $(\kappa(\tf),\lambda(\tf)) = (4,5)$, {\em and} that we extract as much work on average as possible. Work is given by \eqq{eq_work}, noting that $c_0(t) = \kappa(t)$.

This problem therefore has two objectives, which we enforce using an order parameter $\Delta + 10^{-3} \av{W}$. The quantity
\beq
\label{delta}
\Delta = N^{-1} \sum_{i=1}^{N} \left[ \rho(r_i, \tf) - \rho_{\rm ss}(r_i) \right]^2
\eeq
 measures the mean-squared error between the actual profile $\rho(r,\tf)$ and the target profile $\rho_{\rm ss}(r)$, using $N=100$ values of $r_i$. The prefactor $10^{-3}$ assigns slightly greater weight to the steady-state transformation, making this the primary objective, with work extraction as a secondary goal.

Note that the radial distribution functions are normalized such that $\int {\rm d}r\, \rho(r,t) = 1$, i.e., they include the Jacobian factor $r$. This normalization emphasizes the regions where particles are most likely to be found, ensuring that the distribution's weight (and the leading contribution to the order parameter $\Delta$) is concentrated where most of the particles reside.

The default protocol called from \code{load_default_protocol()} immediately sets the control-parameter values to their final values, as shown in \f{fig_abp_1}(a). This protocol fails to effect a state-to-state transformation, as shown in \f{fig_abp_1}(b), generated using \code{visualize_protocol()}, which uses $10^5$ trajectories. The function \code{final_answer} returns \code{order parameter = 0.0115435, Delta = 0.0115435, mean_work = 0}, using $10^6$ trajectories. No work is done because $\kappa$ is not changed.

In practice, we can assign different importance to the two objectives of enacting a state-to-state transformation and extracting work. You might want to experiment with different choices of the order parameter in order to see how the outcome changes as one objective is favored over the other. You could also modify the order parameter to reward efficiency rather than work, accounting for the energy fed to the active particle.

\subsection{Minimizing entropy production during state change in the Ising model}
\label{ising_1}

\begin{figure}[]
   \centering
   \includegraphics[width=\linewidth]{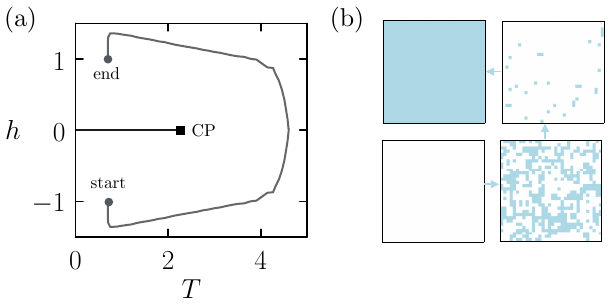} 
   \caption{State change with least entropy production in the Ising model. (a) Protocol that produces least entropy, within the near-equilibrium approximation of \cc{rotskoff2015optimal}. The black line is the line of first-order phase transitions; CP is the Ising model critical point. This protocol is called from \code{default_protocol()}. (b) Time-ordered snapshots of the system under the default protocol, produced by \code{visualize_protocol()}.}
   \label{fig_ising_1}
\end{figure}

The code \code{engine_ising} in the folder \code{ising} simulates state change in the Ising model, a prototype of information erasure and copying in nanomagnetic storage devices. Finding time-dependent protocols that minimize dissipation when performing these processes\c{rotskoff2015optimal,gingrich2016near} is relevant to the problem of reducing energy demands in computation\c{lambson2011exploring}. 

We consider the 2D Ising model\c{binney1992theory} on a square lattice of $N=32^2$ sites, with periodic boundary conditions in both directions. On each site $i$ is a binary spin $S_i = \pm 1$. The lattice possesses an energy function
\beq
\label{ising}
E = -J\sum_{\av{ij}} S_i S_j -h\sum_{i=1}^N S_i,
\eeq
in units such that $k_{\rm B} =1$. Here $J$ (which we set to 1) is the Ising coupling, and $h$ is the magnetic field. The first sum in \eq{ising} runs over all nearest-neighbor bonds, while the second runs over all lattice sites. We begin with all spins down, giving magnetization $m=N^{-1} \sum_{i=1}^N S_i=-1$. 

The model evolves by Glauber Monte Carlo dynamics. At each time step a lattice site $i$ is chosen, and a change $S_i \to -S_i$ proposed. The proposal is accepted with probability 
\beq
P_{\rm Glauber}(\Delta E)=\left( 1+ \exp(\Delta E/T) \right)^{-1},
\eeq
 where $\Delta E$ is the energy change under the proposed move, and is rejected otherwise. Following\cc{engel2023optimal}, lattice sites are chosen sequentially, sweeping through one sublattice followed by the other, rather than at random.

The control parameters of this problem are temperature and magnetic field, $\cee(t) = (T(t), h(t))$. The aim of the problem is to change temperature $T$ and field $h$ from the values $\cee(0) = (0.7,-1)$ to the values $\cee(\tf) = (0.7,1)$, in $t_{\rm f}=100 N$ Monte Carlo steps (i.e. 100 Monte Carlo sweeps). The order parameter to be minimized is the mean entropy production $\av{\sigma}$, where the entropy produced over the course of a simulation is
\beq
\label{ep}
\sigma=E_{\rm f}/T_{\rm f} -E_0/T_0- \sum_{k=1}^{\tf} \Delta E_k/T_k.
\eeq
Here $E_{\rm f}$ and $E_0$ are the final and initial energies of the system, and $\Delta E_k$ and $T_k$ are the energy change and temperature at step $k$ of the simulation. We do not require explicitly that the system's magnetization be reversed, but minimizing entropy production strongly favors magnetization reversal, because this leads to the cancellation of the first two terms in \eq{ep}. 

The function \code{load_default_protocol()} calls the protocol of\cc{rotskoff2015optimal}, which is the optimal protocol within the near-equilibrium approximation of\cc{sivak2012thermodynamic}. This protocol is shown in \f{fig_ising_1}(a), with time as a parameter. This protocol avoids the first-order phase transition line and the critical point. With the default protocol, calling \code{final_answer()} gives an estimate for the entropy production of \code{order parameter} $=86.408 \pm 0.0362$, using 100 samples of $10^3$ trajectories. To appreciate the scale of this result, note that simply switching to the final set of control parameters, without any change of configuration, produces entropy in the amount $2 \times 32^2/0.7 \approx 2.93 \times 10^3$.

The user's task is again to specify the protocol via \code{input_control_parameters.dat}. The code will adjust any value of $T$ below $10^{-3}$ to $10^{-3}$.

One notable feature of this problem is that small perturbations about the simple protocol (the sudden switch to the final set of control parameters) produces little change of microscopic state, because at $T=0.7$ the energy barrier to flipping single spins due to breaking bonds is more than 10 (in units of $\kt$). One way to proceed is to start with a sufficiently large random perturbation\c{whitelam2023demon}; another is to guess a protocol sufficiently close to the optimal one\c{engel2023optimal}.

\section{Worked example: protocol learning using neuroevolution} 
\label{worked_example}

\subsection {Translating an overdamped particle with minimum work}
\label{trap_overdamped_2}

\begin{figure*}[]
   \centering
   \includegraphics[width=0.75\linewidth]{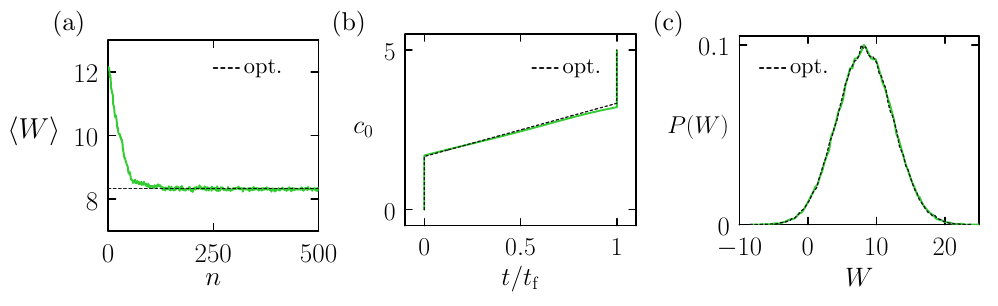} 
   \caption{Worked example: learning an efficient protocol, via neuroevolution, for the translation of an overdamped particle by a harmonic trap~\c{schmiedl2007optimal}; see \s{trap_overdamped_1}. The problem is to specify the time-dependent protocol $c_0(t)$, where $c_0$ is the trap center, in order to minimize the mean work over many realizations of the protocol. Simulation results are shown in green, and results associated with the optimal protocol of \cc{schmiedl2007optimal} are shown in black. (a) Mean work for the highest-ranked individual in the genetic population, as a function of evolutionary time $n$. (b) Protocol learned after several generations, superposed against the optimal protocol. (c) Work distributions associated with the learned and optimal protocols.}
   \label{fig_trap_overdamped_2}
\end{figure*}

In this section we provide a worked example of protocol learning using C++ code. We consider the trap-translation problem of \s{trap_overdamped_1}, and apply to it a neuroevolutionary learning procedure\c{montana1989training,floreano2008neuroevolution,such2017deep}. In this approach, described by the codes \code{ga.cpp} and \code{ga_process.cpp}, the control protocol ${\bm c}(t)$ is expressed as a neural network. A neural network is a flexible functional ansatz that doesn't presuppose any particular functional form, but can (given enough hidden nodes and training time) approximate any smooth function. Some of the features that appear in optimal-control problems are not smooth, and so a neural net must approximate them using rapidly-varying smooth features (see e.g. \s{trap_underdamped_2}).

 The weights of this neural network are trained, by genetic algorithm, to minimize a chosen order parameter. This procedure is relatively simple, and uses, for the most part, old technology. The optimization method is a mutation-only genetic algorithm~\footnote{The reason for the use of mutations only is that we know, for a population of size 2 and a deterministic loss function, that it corresponds in an effective sense to gradient descent in the presence of white noise, and so can be expected to find a good solution if run long enough\c{whitelam2021correspondence}. Other genetic procedures, such as crossover, may be beneficial, but their effective learning dynamics is unknown.}, developed in the 1960s\c{mitchell1998introduction,GA}. The neural network in \code{run_net()} in \code{ga_process.cpp} is a multilayer perceptron, the elements of which have been available since the 1980s\c{lecun2015deep,schmidhuber2015deep}, with the exception of layer normalization, introduced in 2016\c{ba2016layer}. Layer normalization is a useful feature because it ensures that, under simple mutations, the scale of weight-induced signal changes to each neuron does not depend on the number of inputs to each neuron.

The code \code{ga_process.cpp} represents a single member of a genetic population. It encodes the control protocol as a neural network, and calculates the value of the order parameter by applying the control protocol to the system (the neural network is set randomly unless \code{q_read_start_net==1}, in which case it loads from \code{start_net.dat}). To do so it calls functions from the trap \code{engine_trap_overdamped.cpp} via the header
\begin{lstlisting}
#include "engine_trap_overdamped.h"
\end{lstlisting}

The engine code \code{engine_trap_overdamped.cpp} is first turned into a static library, using \code{make library} (ensure that the \code{main()} function is commented out). The \code{ga_process.cpp} code is linked to the resulting library \code{libengine_trap_overdamped.a} by compiling as \code{g++ ga_process.cpp -L. -lengine_trap_overdamped -o sim}. Once linked, \code{ga_process.cpp} mutates the neural network (except for the top-ranked individual in the population, a procedure known as elitism), creates a lookup table specifying the time-dependent protocol, and evaluates the order parameter over $10^4$ trajectories:

\begin{lstlisting}
//mutate net
if((q_production_run==1) && (iteration!=0)){mutate_net();}

//make protocol lookup table
make_lookup_table();

//load protocol
load_protocol();

//visual check of protocol
output_protocol();

//calculate order parameter
op=calculate_order_parameter(10000);
\end{lstlisting}

In the above, \code{make_lookup_table()} converts the output of the neural network into a file \code{input_control_parameters.dat}, which is read by the simulation engine via the command \code{load_protocol()}.

Our goal is to minimize average work, when calculated over 100 samples of $10^4$ trajectories (this calculation is done by \code{final_answer()}). For the purposes of learning we are free to use any number of trajectories; here we run \code{calculate_order_parameter} using $10^4$ trajectories.

The code \code{ga.cpp} runs the genetic algorithm. It monitors, on an HPC cluster, $M=50$ copies of \code{ga_process.cpp} in parallel, each initialized using a randomized neural network. After all have run, it selects the neural networks associated with the $M_0=10$ best (smallest) values of the order parameter \code{op}, and constructs another population of $M$ neural networks by cloning and mutating the $M_0$ parent networks. Mutations involve a Gaussian random number added to each neural-network parameter. This process continues until terminated. The genetic algorithm can be run on a SLURM cluster using \code{g++ -Wall -o deschamps ga.cpp -lm -O} and \code{sbatch deschamps.sh}, replacing the queue commands in the \code{deschamps.sh} script (deschamps is the ``water carrier'') with those appropriate for your HPC cluster.

The results of this learning process are shown in \f{fig_trap_overdamped_2}. Panel (a) shows the value of \code{calculate_order_parameter(10000)} for the best-performing member of the population of 50, as a function of generation number $n$. Panel (b) shows the protocol \code{output_protocol()} learned at the end of the process, which is similar to that of the optimal protocol shown in \f{fig_erasure_1}(a). It is not exactly the same, but its performance and that of the optimal protocol are difficult to distinguish when evaluated over a finite number of trajectories. For the learned protocol, the average \code{final_answer()} returns the estimate \code{order parameter} $= 8.33697 \pm 0.00439$, which is equal, within statistical error, to the optimal mean work value of $25/3$. Panel (c) shows the work distributions, generated by \code{visualize_protocol()}, for the learned and optimal protocols, which are essentially identical.

The folder \code{trap_overdamped} contains both the simulation engine \code{engine_trap_overdamped.cpp}, and the code to train a control protocol via neuroevolution, \code{ga.cpp} and \code{ga_process.cpp}. It also includes the protocol learned in this way, \code{input_control_parameters_learned.dat}. This can be called via \code{load_protocol()}, after changing the name of that file to \code{input_control_parameters.dat}.

In the following sections we briefly present the protocols learned by neuroevolution for the other problems in this set. These protocols can be found in the relevant folders, in the files \code{input_control_parameters_learned.dat}.

\subsubsection{Python implementation of the genetic algorithm}

The python version of \code{trap_overdamped} contains a code, \code{ga.py}, that runs a similar neuroevolutionary optimization procedure using the Python versions of the engines. It does this locally on a single CPU or GPU. The neural network is implemented using PyTorch in the file \code{nets.py}. $M$ copies of this neural network are created and randomly initialized. The corresponding protocols are evaluated, with all trajectories for each protocol running in parallel (further parallelization could be achieved by distributing the neural-network protocol evaluations over multiple CPUs or GPUs). Parameters such as the population size, mutation rate, and neural network architecture can be changed in the \code{config} dictionary. The genetic algorithm runs for \code{num_generations} generations. It periodically creates plots of the best-performing protocol and corresponding work distribution, together with a movie showing the effect of the protocol on the particle distribution. 

\subsubsection{JAX implementation of automatic differentiation}

Also included in the benchmark package is an example of an alternative means of computing optimal thermodynamic protocols using automatic differentiation, as first described in Reference~\cite{engel2023optimal}. The script \texttt{optimize\_overdamped\_brownian.py} implements the minimum work translation of an overdamped particle as described above, but rather than using a genetic algorithm to arrive at the minimum work protocol, the gradient of the work with respect to parameters describing the protocol is computed with \texttt{jax.grad()} and used to arrive at the optimal protocol parameters via stochastic gradient descent. In the example code, the protocol is parameterized using 8 discrete points joined by linear interpolations, and discrete jumps at the beginning and end of the protocol are permitted. Other parameterizations, including neural networks, are possible and should be straightforward to implement using the \texttt{flax} library. Brownian dynamics are implemented using the Jax-MD package using a simple first order integration scheme~\cite{JAXMDSchoenholz_2021} and mapped over a batch of 5000 particles using \texttt{jax.vmap}. The gradient over the batch is computed with \texttt{jax.value\_and\_grad} and used with the \texttt{Adam} optimizer from the \texttt{optax} library (other optimizers can easily be experimented with). Running \texttt{optimize\_overdamped\_brownian.py} from the command line will perform 100 iterations of gradient descent with \texttt{Adam} and plot the work as a function of iteration and the evolution of protocols and generate an animation, \texttt{particle\_histogram.gif}, showing the final protocol dragging the distribution of Brownian particles. The automatic differentiation approach is able to recover the analytical solution of minimum work, 25/3 for $\tf = 1$, and can do so rapidly, taking a few minutes on a GPU. 

\subsection {Translating an underdamped particle with minimum work}
\label{trap_underdamped_2}

\begin{figure*}[]
   \centering
   \includegraphics[width=0.75\linewidth]{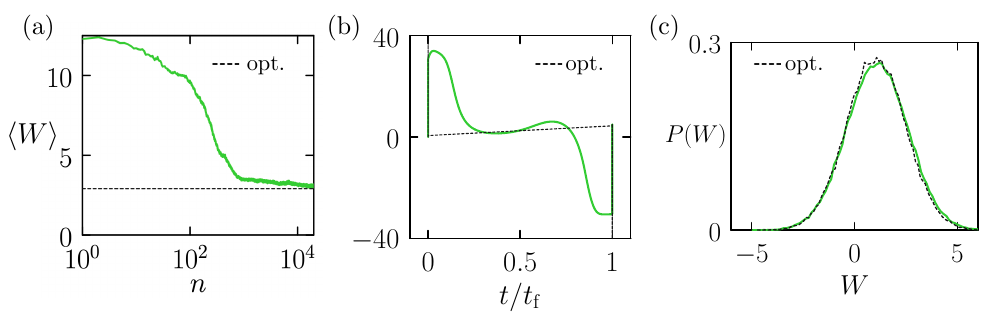} 
   \caption{Learning an efficient protocol, via neuroevolution, for the translation of an underdamped particle by a harmonic trap~\c{gomez2008optimal}; see \s{trap_underdamped_1}. The problem is to specify the time-dependent protocol $c_0(t)$, where $c_0$ is the trap center, in order to minimize the mean work done upon many translations. Simulation results are shown in green, and results associated with the optimal protocol of \cc{gomez2008optimal} are shown in black. (a) Mean work for the highest-ranked individual in the genetic population, as a function of evolutionary time $n$. (b) Protocol learned after several generations, superposed against the optimal protocol. (c) Work distributions associated with the learned and optimal protocols.}
   \label{fig_trap_underdamped_2}
\end{figure*}

The neuroevolution procedure applied to the problem of \s{trap_underdamped_1} results in the protocol shown in \f{fig_trap_underdamped_2}. Because of the need for sharp impulses to control the velocity of the particle, this problem is harder than its overdamped counterpart. The learning algorithm finds a protocol that is visually quite different to the optimal protocol \eq{opt}, displaying rapidly-varying smooth features that act as regularized versions of the delta-function impulses at the start and end of the protocol. With the integration \code{timestep=0.01}, chosen so that the mean work is invariant to a reduction of the timestep by a factor of 10, the function \code{final_answer()} returns a value for the order parameter (mean work) of $3.13804 \pm 0.00283$, about 8\% larger than the optimal value. The work distributions associated with the optimal and learned protocols, shown in \f{fig_trap_underdamped_2}(c), are similar, but can be distinguished.

Neural networks require several training steps in order to learn high-frequency features, such as are needed to exert efficient control over an underdamped particle. Faster convergence might be obtained by using an ansatz that contains explicit spike-like features, if you allow yourself some knowledge of the optimal protocol and attempt to simply find a numerical approximation to it. 

Beware the ability of learning algorithms to exploit numerical artifacts: some runs of the genetic algorithm produced protocols with spiked features large enough to break the integrator, producing unphysical mean work values well below the true optimum. If this danger exists it is necessary to check that reducing the integration step does not change the outcome of the simulation (this is also a good idea in general). Genetic algorithms have exploited numerical bugs in other contexts in order to achieve an objective by subterfuge\c{lehman2020surprising}.

One strategy that can be used here is to run an initial learning phase using an illegally large integration timestep, in order to make rapid progress, before implementing a safe timestep.

\subsection{Maximizing the probability of logic erasure using an underdamped particle}
\label{erasure_2}

\begin{figure*}[]
   \centering
   \includegraphics[width=\linewidth]{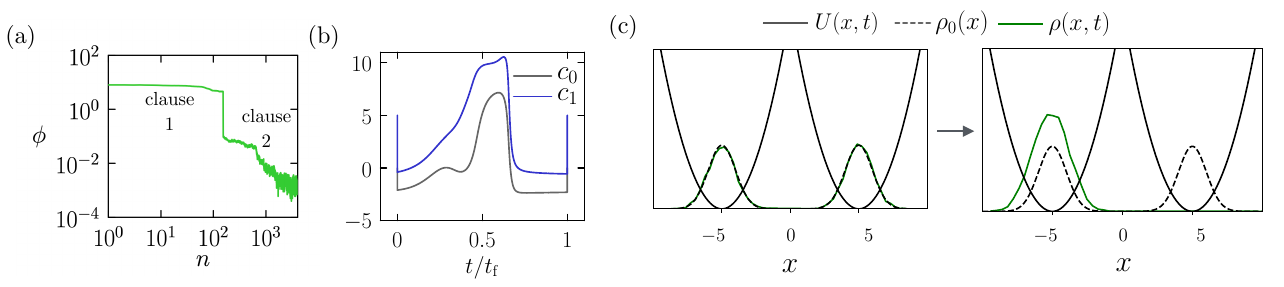} 
   \caption{Learning an efficient protocol, via neuroevolution, for the erasure procedure of \s{erasure_1}. (a) Value of \code{custom_order_parameter()} as a function of the number of evolutionary generations $n$; the switch to the standard order parameter (the mean failure rate) occurs after about 100 generations. (b) The protocol obtained after training. (c) Snapshots of the system under the influence of the protocol. Compare the effect of the default protocol, shown in \f{fig_erasure_1}. Here $U(x,c_0(t))$ is the potential, $\rho_0(x)$ is the associated Boltzmann distribution, and $\rho(x,t)$ is the distribution of particle positions.}
   \label{fig_erasure_2}
\end{figure*}

\begin{figure*}[]
   \centering
   \includegraphics[width=0.8\linewidth]{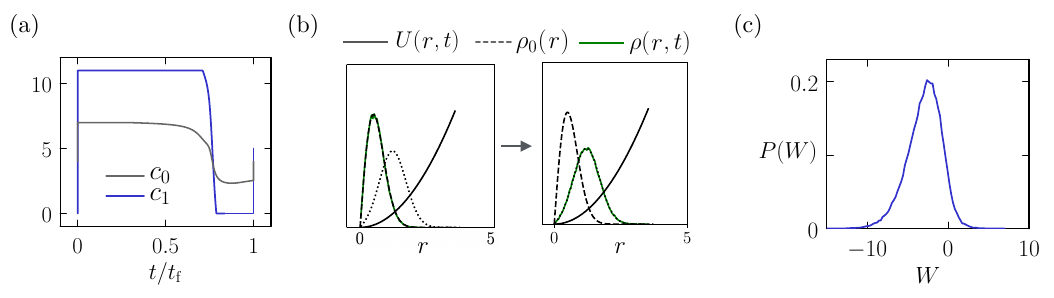} 
   \caption {Learning an efficient protocol for an Brownian particle in a trap, modified from\ccc{baldovin2023control,casert2024learning}. The goal of this problem is to induce a state-to-state transformation while extracting as much work as possible. (a) The learned protocol for the two control parameters ${\bm c(t)} = (\kappa(t),\lambda(t))$. (b) Time-ordered snapshots created by \code{visualize_protocol()}, illustrating the action of the protocol. Here $U(r,{\bm c}(t))$ is the harmonic potential in which the particle is confined, $\rho_0(r)$ is the associated Boltzmann distribution, and $\rho(r,t)$ is the distribution of particle positions. The target distribution is the outer dotted profile: a state-to-state transformation is enacted. (c) Work distribution resulting from the protocol. On average, this protocol extracts about 2.9 $\kt$ of energy from the system.}
   \label{fig_abp_2}
\end{figure*}

The neuroevolution procedure applied to the problem of \s{erasure_1} results in the protocol shown in \f{fig_erasure_2}. To facilitate learning, we replaced the standard order parameter (the erasure error rate) with \code{custom_order_parameter()}, defined in \code{engine_erasure.cpp}: this is a two-clause function that returns the standard order parameter if the erasure probability exceeds 0.9, and otherwise returns a function of the mean position of the particle at the end of the trajectory. This tactic is sometimes called curriculum learning. We start from a null protocol, from which the erasure probability remains close to 50\% for small initial changes of the neural network, providing little signal for the learning algorithm to use. The second clause of the custom order parameter teaches the algorithm to move the particle closer to the desired potential well, before the standard order parameter switches on. An alternative would be to start from a protocol that produces a reasonable erasure probability and use the regular order parameter throughout. However, note that for the system parameters considered here, the default protocol described in \s{erasure_1} produces an erasure probability of only 53\%.

The learned protocol shown in \f{fig_erasure_2}(b) is much more efficient than the default protocol: running \code{final_answer()} gives the value of the order parameter, using 100 samples of $10^4$ trajectories, as $0.001446 \pm 3.57329 \times 10^{-5}$, corresponding to about 14.5 failures per $10^4$ erasures. Two time-ordered snapshots are shown in panel (c), using \code{visualize_protocol()}. The movie \code{output_movie.mp4} generated by the same routine is more instructive, showing parts of the protocol to involve a single potential well jumping back and forth, either side of the particle-position distribution, applying kicks in an effort to ensure reliable erasure.

\subsection {Enacting a state-to-state transformation and extracting work from an active Brownian particle in a trap}
\label{abp_2}

\f{fig_abp_2}(a) shows the protocol learned by neuroevolution when instructed to enact a state-to-state transformation and to extract as much work as possible from the active Brownian particle in a trap. For this protocol, the function \code{final_answer()} returns the results \code{order parameter = -0.00295971, Delta = 5.10897e-06, mean_work =-2.96482}: the state-to-state transformation is enacted to high precision, extracting almost $3 \kt$ units of work in the process. An in\cc{casert2024learning}, the work extraction protocol involves an initial input of work, as the trap parameter $\kappa$ is increased, with the extraction of work happening later in the protocol, when $\kappa$ is returned to its resting value.

\subsection{Minimizing entropy production during state change in the Ising model}
\label{ising_2}

\f{fig_ising_2} shows the protocol learned by neuroevolution when instructed to change state in the Ising model while minimizing entropy production, overlaid on the default protocol (the near-equilibrium protocol of\cc{rotskoff2015optimal}). The learned protocol is markedly asymmetric, a feature shared by the protocol of\cc{engel2023optimal}, which was identified by gradient descent. The function \code{final_answer()} shows the learned protocol to produce entropy in the amount \code{order parameter} $= 52.6909 \pm 0.030566$, about 61\% of that produced by the near-equilibrium approximation. 

Here the learned protocol is constrained to be smooth at initial and final times\c{whitelam2023demon}, an aesthetic choice motivated by the idea that abrupt jumps of temperature (in a real system) can drive a thermal bath out of equilibrium. You could experiment to see if the jumps that appear in other problems are beneficial here. 

For this problem there many protocols that look visually quite different, but produce similar values of entropy\c{gingrich2016near}.

\begin{figure}[]
   \centering
   \includegraphics[width=\linewidth]{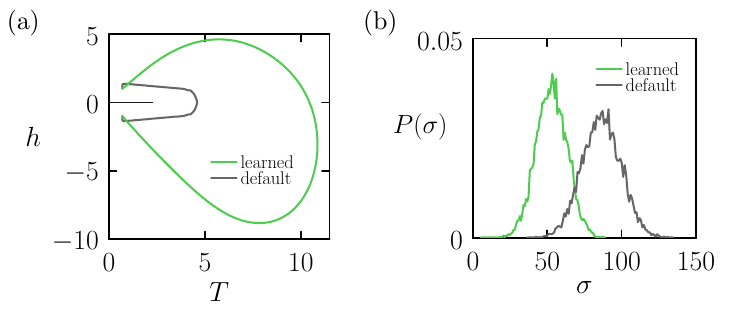} 
   \caption{Learning an efficient protocol for state change in the Ising model with least entropy production. (a) Learned protocol (green), compared with the default protocol (grey). The latter is the optimal protocol within the near-equilibrium approximation of \cc{rotskoff2015optimal}. (b) Entropy production histograms for the two protocols, produced by \code{visualize_protocol()}.}
   \label{fig_ising_2}
\end{figure}

Table~\ref{tab2} summarizes the results of the neuroevolutionary algorithm applied to the set of benchmark problems. For this algorithm, as for other numerical algorithms, there is no guarantee of optimality, and we must therefore be pragmatic, checking that similar results are obtained from different runs of the algorithm (including runs with different hyperparameter choices) and different initial protocols. 

\subsection{A note on genetic algorithms}

Genetic algorithms (GAs) have been around since the 1960s\c{mitchell1998introduction,GA}, and have been used to train neural networks since at least the 1980s\c{montana1989training}. The approach described in \s{worked_example} is inspired by neuroevolutionary methods used in 2017 to learn control strategies for Atari games\c{such2017deep}, with some small differences of implementation. 
\begin{table*}[]
\renewcommand{\arraystretch}{1.3} 
\centering
\begin{tabular}{|C{3.2cm}|C{4.2cm}|C{3.6cm}|C{3cm}|C{2cm}|}
\hline
\textbf{Problem} & \textbf{Description} & \textbf{Objective} & \textbf{\code{final_answer()}} & \textbf{Optimal value} \\
\hline
\code{trap_overdamped} & Overdamped Langevin particle in a moving harmonic trap & Minimize work & $8.33697 \pm 0.00439$ & $25/3$ \\
\code{trap_underdamped} & Underdamped Langevin particle in a moving harmonic trap & Minimize work & $3.13804 \pm 0.00283$ &2.91 \\
\texttt{erasure} & Logic erasure via underdamped dynamics & Minimize error rate &  $1.446 \times 10^{-3} \pm 3.57329 \times 10^{-5}$ &Unknown \\
\texttt{abp} & Active Brownian particle in a harmonic trap & Enact state-to-state transformation and maximize extracted work & $-3.0 \times 10^{-3}$ ($\Delta = 5.1 \times 10^{-6}$, $\av{W} = -2.9$)& Unknown \\
\texttt{ising} & 2D Ising model undergoing state change & Minimize entropy production & $52.6909 \pm 0.030566$ & Unknown \\
\hline
\end{tabular}
\caption{The results of neuroevolution applied to the benchmark systems, using the C++ code \code{ga.cpp} in the folder \code{trap_overdamped}. A Python version of the genetic algorithm routine can be found in the corresponding Python folder, while the corresponding JAX folder contains code to optimize the overdamped trap protocol by automatic differentiation.}
\label{tab2}
\end{table*}

GAs can handle large numbers of parameters (e.g. tens of millions in\ccc{such2017deep,whitelam2022training}, which is large by the standards of statistical mechanics, at least!), and, where comparison exists, they perform as well as (reach similar loss values to) gradient-based methods\c{such2017deep,whitelam2022training,whitelam2023demon}. There are theoretical reasons to expect this similarity. The zero-temperature Monte Carlo algorithm, which is a mutation-only genetic algorithm with a population of size 2, is for small mutations equivalent to noisy gradient descent on the loss surface\c{whitelam2021correspondence}. So, while very different in implementation, mutation-only GAs do effectively what gradient descent does.

The strengths of GAs are simplicity and robustness. Unlike gradient-based methods, genetic algorithms do not require knowledge of the gradient of the objective function. They explore weight space by biased stochastic diffusion, and can locate a signal even if the landscape is sparse, noisy, or non-differentiable\c{whitelam2022training}. GAs are therefore suited to problems for which good initial guesses are unavailable, or where the order parameter is defined only at the end of a trajectory. In physics, where we often care about a single scalar quantity such as average work, entropy production, or erasure rate, this compatibility with sparse rewards is a natural advantage. GAs also handle feedback-control protocols with no greater conceptual complexity than feedforward (time-only) protocols\c{whitelam2023demon,whitelam2023train}, aside from the need to include additional input nodes in the neural network to represent the system state.

The weakness of GAs is their lack of speed. Gradient-based methods can take large directed steps when gradients can be calculated, while GAs follow the gradient by biased diffusion. As a result, GAs require many evaluations, and are slower than gradient-based methods.

When choosing between GAs and gradient-free optimization methods in physics, the key question to ask is the following: how difficult is it, in terms of code design, to obtain reliable gradients for your system? If reliable gradients can be calculated, gradient-based methods are the natural choice. If gradients are undefined, noisy, or memory-intensive to compute, then GAs offer a relatively simple alternative with similar learning capacity. Indeed, combining the two approaches could be a useful strategy.

\section{Conclusions}

We have presented a set of computer codes designed to test methods for optimizing time-dependent control protocols in fluctuating nonequilibrium systems. The problem set, available at the GitHub repository \href{https://github.com/protocol-benchmarks/NESTbench25}{\texttt{protocol-benchmarks/NESTbench25}}, contains five models, each with an associated optimization objective, coded in C++ and Python. These benchmark systems are simple enough to run on a laptop, but challenging enough to test the capabilities of modern optimization methods. We have included with the set a worked example and a set of results derived from neuroevolutionary learning, and we have discussed some of the challenges associated with each problem.

Benchmarks play an important role in the machine-learning literature by providing standardized test problems. Our benchmark set is designed to provide a set of standardized tests for physical scientists working on control problems. Our hope is that the set is instructive, and allows users to develop understanding of the strengths and weaknesses of particular methods of optimization.  

\section{Acknowledgements}

The motivation for this work came in part from discussions at the NSF Workshop on the Future of AI in the Mathematical and Physical Sciences at MIT in March 2025, and S.W. thanks the organizers and participants of that meeting. We are grateful to Grant Rotskoff for providing the default protocol, taken from \cc{rotskoff2015optimal}, for the Ising-model problem. Our code can be found at the GitHub repository \href{https://github.com/protocol-benchmarks/NESTbench25}{\texttt{protocol-benchmarks/NESTbench25}}. S.W. performed work (and emitted heat) at the Molecular Foundry at Lawrence Berkeley National Laboratory, supported by the Office of Basic Energy Sciences of the U.S. Department of Energy under Contract No.~DE-AC02-05CH11231. This research used resources of the National Energy Research Scientific Computing Center (NERSC), a U.S. Department of Energy Office of Science User Facility. M.C.E. gratefully acknowledges Natural Sciences and Engineering Research Council of Canada (NSERC) Discovery Grant RGPIN-2024-06144.


%

\end{document}